\documentclass[prb,twocolumn,showpacs]{revtex4-1}

\usepackage{graphicx}
\usepackage{dcolumn}
\usepackage{bm}
\usepackage{amsmath}
\usepackage{epstopdf}
\usepackage{wasysym}

\begin{document}
\title{Collective dynamical skyrmions excitations in magnonic crystal}
\author{M.~Mruczkiewicz} \email{m.mru@amu.edu.pl}
\author{P.~Gruszecki}
\author{M.~Zelent}
\author{M.~Krawczyk} \email{krawczyk@amu.edu.pl}
\affiliation{Faculty of Physics, Adam Mickiewicz University in Poznan, Umultowska 85, Pozna\'{n}, 61-614, Poland}
\date{\today}

\begin{abstract}

We investigate theoretically skyrmion magnonic crystal, i.e., the dynamics of the magnetization in a chain of the ferromagnetic nanodots being in skyrmion magnetic configuration. We show that collective excitations are possible to be observed in the structure. We present the dispersion relation of the coupled skyrmions. It exhibit a periodical property in dependence on wave vector, characteristic feature of the band structure in magnonic crystals. Spatial analysis of the magnetization amplitude associated with the magnonic bands confirms type of the excited modes, as breathing and clockwise gyrotropic dynamical skyrmions. These high and low frequency excitations, propagate with negative and positive group velocity, respectively, and can be explored for study fundamental properties and technological applications in spintronics and magnonics.

\end{abstract}

\maketitle

Properties of the magnetic structure are determined by set of numerous parameters, like a saturation magnetization, exchange constant,  anisotropy, damping factor or geometry. Because of this large degree of freedom in manipulating of the magnetic structures, the physics of magnetic materials is extremely vast, rich in phenomena and useful for application in microwave technology, information storing and processing. Another parameter which has important impact on properties of the magnetization arrangement and its dynamics is Dzyaloshinskii-Moriya interaction (DMI). The DMI is also called an antisymmetric exchange interaction and it appears as a result of spin-orbit coupling in non-centrosymmetric magnets  or it comes from spin-orbit interactions at the interface of ultrathin magnetic films.\cite{117,118}

Although influence of the DMI on magnetization texture at room temperature has not yet been demonstrated, its influence on dynamical properties is significant. Recently spin waves (SWs) propagating in thin films have been investigated analytically,\cite{45} numerically\cite{47,48} and also experimentally,\cite{91} showing the great importance of DMI to nonreciprocal character of SW propagation and electric field controllable phase shift of SWs. In addition, DMI can destroy the degeneracy of chitrality of the vortex state.\cite{115}

At low fields and temperatures presence of DMI leads to appearing of the exotic spin textures. As result of simultaneous contribution with exchange interactions, various arrangements of magnetization can spontaneously realize, e.g., helical\cite{77}  or conical\cite{21} spin textures. An interesting phenomena recently discovered in magnetic structures is a skyrmion.\cite{0_TS} It was shown with Monte Carlo and micromagnetic simulations that single skyrmion can nucleate in the isolated disk\cite{44,65,66} or skyrmionic spin texture can appear in ultrathin film, i.e., skyrmionic crystal (SkX). The first experimental observation of SkX was made with the use of neutron scattering\cite{85} after the theoretical predictions.\cite{Bogdanov}  Analytic, numerical and experimental investigations are conducted at the present moment in order to understand the formation and properties of SkX lattices. Material properties are optimized for stabilization of SkX phase at large range of temperature and magnetic field. Positive influence of anisotropy on stabilization was also showed.\cite{64} It was found that dimension plays important role in stabilization of the SkX: in two-dimensional ultrathin films can exhibit the SkX texture at wide ranges of temperatures and fields showing even near room-temperature formation of the skyrmions array.\cite{20} Structuralized and confinement structures are also studied,\cite{71} they not only permit on designing manually SkX lattices, but extends the temperature stability range. Since the discovery of skyrmions in magnetic structures, they are considered for potential application as a part of the memory devices,\cite{kiselev2011chiral} magnetic racetracks\cite{5} and logic devices.\cite{zhang2014magnetic} 
This is due to their high mobility under applied currents. The current needed to control skyrmion can be few orders of magnitude lower than for control of the domain wall.\cite{66} 

\begin{figure}[!ht]
\includegraphics[width=0.45\textwidth]{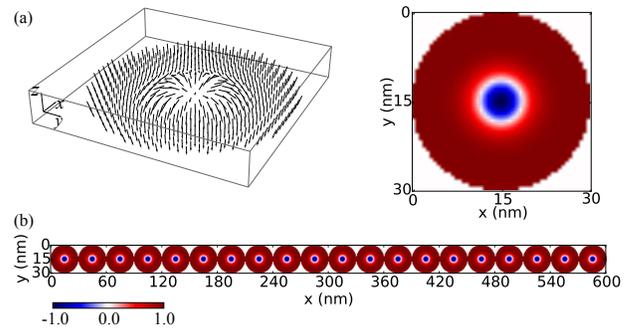}
\caption{(Color online) A structure under consideration. The bias magnetic field $B_{\text{ext}}$ is out-of-plane plane and directed along the $z$ axis. (a) The 3D plot shows the magnetic configuration in a single dot (left image) and the $m_{z}(t_{0})$ component of the magnetization (right). (b) The $m_{z}(t_{0})$ component of the magnetization is plotted as function of $x$ and $y$ for the array of 20 nanodisks with 30 nm diameter. }
\label{structure}
\end{figure}

The dynamical excitations of skyrmions and SkX started to be investigated, too. They can be induced with external field, spin waves, electric current or temperature. An examples of conducted studies is interaction with magnetic droplet\cite{55} and SWs scattering on single skyrmion.\cite{30} The appearance of the resonant magnetization oscillations in a single skyrmion under influence of the external dynamic magnetic field was presented, i.e., breathing,\cite{30,44,55,56,57} clockwise and counter-clockwise modes.\cite{56,57} Also the melting of SkX lattice under external dynamic magnetic field was shown.\cite{38} The interaction of exotic spin texture with SWs are also studied since it was suggested that magnons can be used to control the motion of skyrmions.\cite{26} The dynamics of coupled skyrmions\cite{wang2014microwave} and collective excitation\cite{58} were considered as well. The  spintronic application of skyrmion-based  spin transfer nano-oscillator (STNO) was proposed \cite{zhang2015current}.

In this Letter, we numerically study collective skyrmion dynamics in the chain of the magnetic nanodots with skyrmion texture. Investigated system is analogical to one presented in Ref.~[\onlinecite{99}], where collective excitations of the vortex states were studied. However, in our study the dots are in skyrmion state and have almost of two orders smaller diameter. We show, that collective dynamical skyrmions appear in this magnonic crystals due to dipole coupling between nanodots.  Moreover, we demonstrate, that the dynamic signal can be transmitted through the chain by different bands, which are connected with basic excitations of the isolated skyrmion: breathing and gyrotropic type of modes. These high and low frequency bands are characterized by positive and negative group velocities.

\begin{figure}[!ht]
\includegraphics[width=0.45\textwidth]{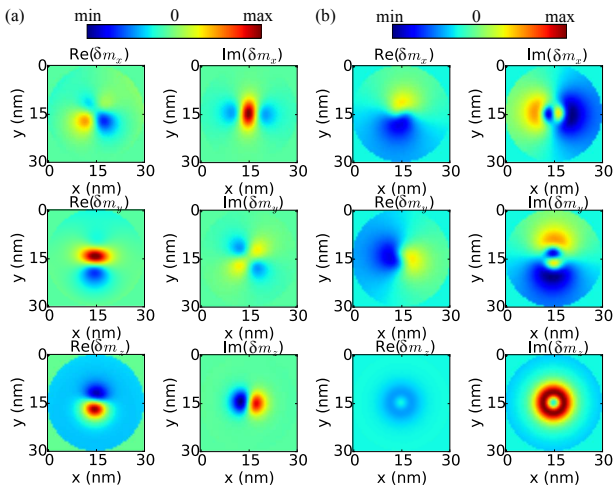}
\caption{(Color online) Dynamical components of the magnetization $\delta m_{x}$, $\delta m_{y}$ and $\delta m_{z}$ of the clockwise, 0.98 GHz (a) and breathing, 13.24 GHz (b) modes obtained using FDTD. In the left (right) column the real (imaginary) part of $\delta m$ is shown.}
\label{single}
\end{figure}

We have performed finite-difference time-domain  (FDTD) simulations with mumax$^3$ solver\cite{4899186} and we have used uniformly discretized grid with the size of the cell $0.6 \times 0.6 \times 1.0$ nm$^3$. In mumax$^3$, the time derivative of {\bf M} is defined as the torque ${\bf \tau}$: $
\frac{\partial {\bf M}({\bf r},t)}{\partial t}={\bf \tau}$. The basic micromagnetic properties are described with Landau-Lifshitz torque:
\begin{equation}
{\bf \tau}= \gamma \frac{1}{1+\alpha^2}\left({\bf M} \times {\bf B_{\mathrm{eff}}}+\alpha \left({\bf M} \times \left({\bf M} \times {\bf B_{\mathrm{eff}}} \right) \right) \right),
\end{equation}
where  $\alpha$ is the dimensionless damping parameter and ${\bf B_{\mathrm{eff}}}$ is the effective magnetic induction field, which can have contributions from: external magnetic field, magnetostatic demagnetizing field, Heisenberg exchange field, Dzyaloshinskii-Moriya exchange field, magneto-crystalline anisotropy field and thermal field. For purpose of our study, following contributions to the effective field ${\bf B_{\mathrm{eff}}}$ are taken into account:
\begin{equation}
{\bf B_{\mathrm{eff}}}={\bf B_{\mathrm{ext}}}+{\bf B_{\mathrm{demag}}}+{\bf B_{\mathrm{exch}}}+{\bf B_{\mathrm{DM}}}+{\bf B_{\mathrm{anis}}}, 
\end{equation}
where ${\bf B_{\mathrm{ext}}}$, ${\bf B_{\mathrm{demag}}}$, ${\bf B_{\mathrm{exch}}}$, ${\bf B_{\mathrm{DM}}}$, ${\bf B_{\mathrm{anis}}}$ are external magnetic field, demagnetizing, exchange, Dzyaloshinskii-Moriya and anisotropy fields, respectively.\cite{4899186}
\begin{figure}[!ht]
\includegraphics[width=0.45\textwidth]{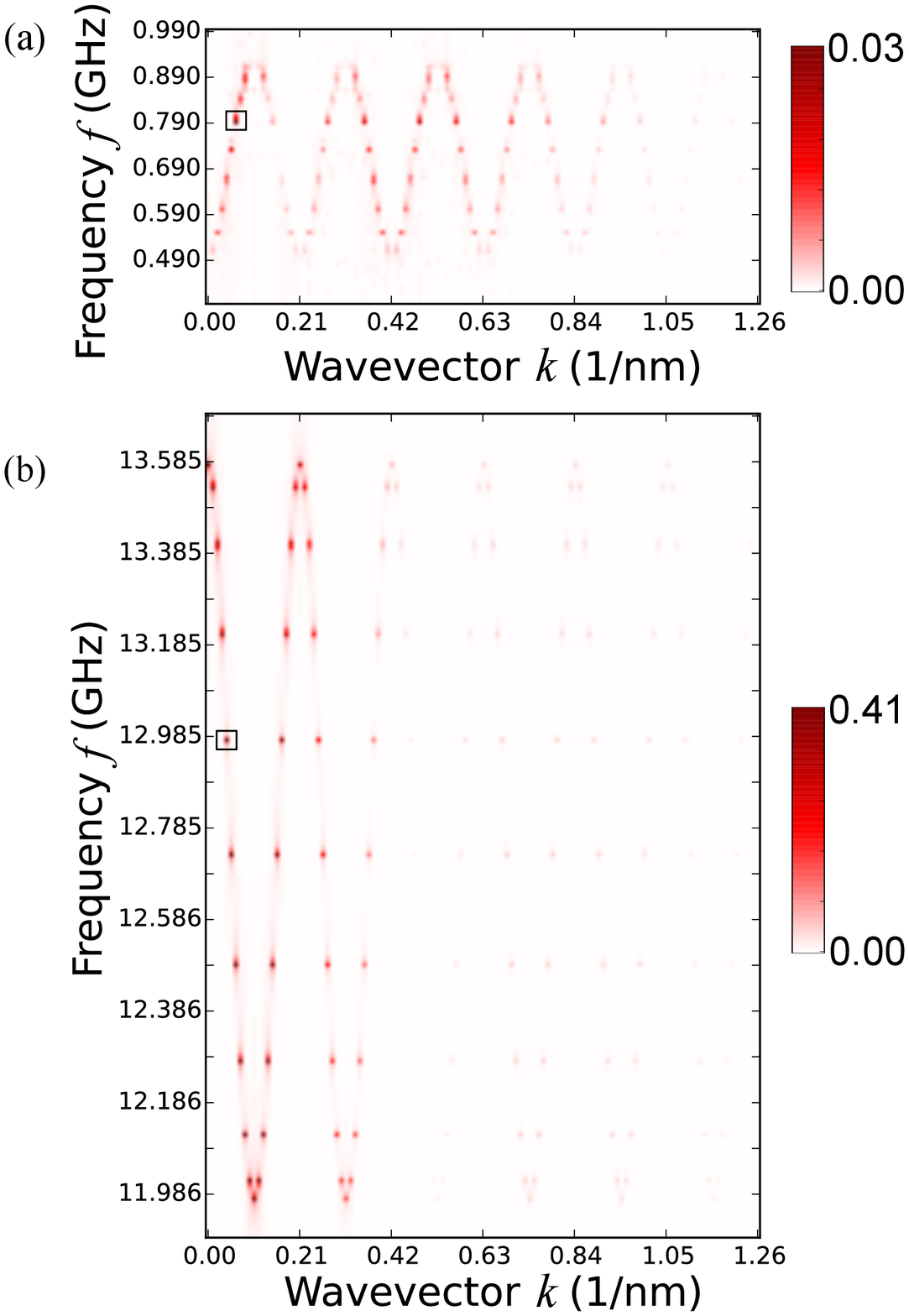}
\caption{(Color online) The $\left| \mathrm{2DFT} \right|$ function showing (a) the dispersion relation of the collective clockwise gyrotropic skyrmion excitation and (b) the dispersion relation of the collective breathing skyrmion excitation. The black squares indicate wavenumbers that are investigated in manuscript. The $\left| \mathrm{2DFT} \right|$ is normalized to maximum value for each graph seperately. The siganl in (a) is $\approx$ 10 times weaker then in (b).}
\label{disp}
\end{figure}

We have used material parameters of the system with perpendicular anisotropy that correspond closely to Pt/Co structure and have been used in the Refs.~[\onlinecite{44,48}], apart from diameter of the nanodisks and damping factor. These two parameters were changed for enhancement of the dispersive character of collective skyrmion excitation and better resolution. The parameters that were used are: thickness of the dots $d=1$ nm, magnetization saturation $M_{\mathrm{S}}= 10^6$ A/m, exchange constant $A_{\mathrm{ex}}= 1.5 \times 10^{-11}$ J/m, uniaxial anisotropy constant $K_{\mathrm{u1}}= 10^6$ J/m$^3$, damping constant $\alpha=0.0001$ and DMI constant $D_{\mathrm{ind}}=0.003$ J/m$^2$. We keep in calculations the out of plane external magnetic field fixed to $B_{\mathrm{z}}=0.1$ T. The micromagnetic simulations were performed in two stages. In the first step the magnetization configuration of initial state was postulated with the N\'eel skyrmion in the center of each dot. The system was relaxed to achieve energy minimum.\footnote{We have used relax() function in mumax$^3$} Due to that procedure the stabilized magnetization vector distribution ($m_{x}(x,y,t_{0})$, $m_{y}(x,y,t_{0})$, $m_{z}(x,y,t_{0})$) was obtained.


\begin{figure}[!ht]
\includegraphics[width=0.45\textwidth]{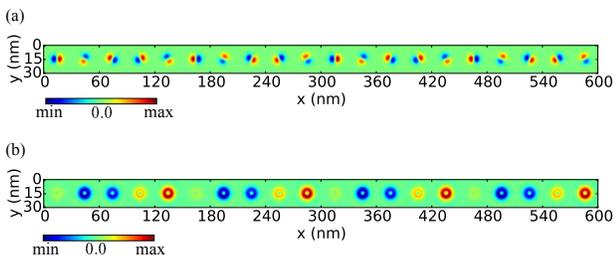}
\caption{(Color online)(a) Real part of $\delta m_{z}$, $\mathrm{Re}(\delta m_{z})$ of the collective clockwise gyrotropic mode with wavenumber $k=0.6 \times \frac{\pi}{a}$. (b) $\mathrm{Re}(\delta m_{z})$ of the collective breathing mode with wavenumber $k=0.4 \times \frac{\pi}{a}$.}
\label{profiles}
\end{figure}

In the second stage the simulations of the magnetization dynamics were performed. The magnetization distribution obtained in the first stage were influenced by an alternating external magnetic field $h_{\mathrm{dyn}}$ that was directed either out of plane ($z$) or in-plane (along $x$ axis). This dynamic field $h_{\mathrm{dyn}} = h_{\mathrm{amp}}\mathrm{sinc}[2\pi f_{\mathrm{signal}} (t-t_{\mathrm{shift}})]$ is characterized with cutoff frequency $f_{\mathrm{signal}}=16$ GHz and is shifted in time by $t_{\mathrm{shift}}= 10^{-9}$ s. The amplitude paramter of the sinc signal, $h_{\mathrm{amp}}$ was choosen so that the maximum absolute value of dynamic magnetic field was $0.01$. The excitation field $h_{\mathrm{dyn}}$ is uniform and limited to the dot. In dynamic simulations, we used the Dormand-Prince method with adaptive time step control for advancing the Landau-Lifshitz equation. The maximum time step was defined to $10^{-12}$ s and minimum time step to $10^{-15}$ s. The total time of the simulations was limited to $3 \times 10^{-7}$ s and dynamic magnetization components were saved with sampling interval  $3 \times 10^{-11}$ s. Such specification allows for calculating Fourier transform to frequency domain with resolution below 10 MHz within the  range slightly exceeding  $(0,f_{\mathrm{signal}})$.


The first  structure under investigation is isolated disk with thickness $d=1$ nm and diameter $2R =30$ nm. Dynamics of skyrmions in isolated disk was already investigated and our simulations are in agreement with these results.\cite{44} The ground state of the isolated disk is presented in Fig.~\ref{structure}(a). We find two intensive skyrmion modes below cutoff frequency of the signal:  $m_{\mathrm{CW}}$ clockwise gyrotropic mode at 0.98 GHz (excited with in-plane external magnetic field) and $m_{\mathrm{Br}}$ breathing mode  at 13.24 GHz (excited with out-of-plane external magnetic field).\footnote{There counter-clockwise gyrotropic mode also exists in the investigated structures, however at frequency 30.19 GHz, well above $f_{\text{signal}}$ as was verified by independent simulations.} The discrete-time Fourier transform (DTFT) was performed for magnetization components, separately at each point of the 2D FDTD grid and value of DTFT at frequency of interest was plotted as function of $x$ and $y$. The function obtained in this way shows dynamic amplitude distribution (its real and imaginary parts, $\mathrm{Re}(\delta m)$ and $\mathrm{Im}(\delta m)$, respectively) as presented in Fig.~\ref{single}.  The character of the breathing mode is clear to interpret, since it shows that largest changes of the magnetization $z$ component in form of a ring shape (Fig.~\ref{single}(b)), meaning that skyrmion is dynamically expanding and compressing. The character of the clockwise mode is more difficult to intuitively interpret from profiles. However, it can be visualized dynamically according to the formula \cite{mamica2011effect, mamica2014effects}:
\begin{equation}
 \begin{split}
 m_{i}(x,y,t) = 
 m_{i}(x,y,t_{0}) + \mathrm{cos}(2 \pi t)\mathrm{Re}(\delta m_{i}) + \\ \mathrm{sin}(2 \pi t)\mathrm{Im}(\delta m_{i}),
 \end{split}
 \end{equation}
where index $i$ indicates the $x$, $y$ and $z$ components. The animation showing time evolution of all three modes can be found in the supplementary material.\cite{animation}

\begin{figure}[!ht]
\includegraphics[width=0.48\textwidth]{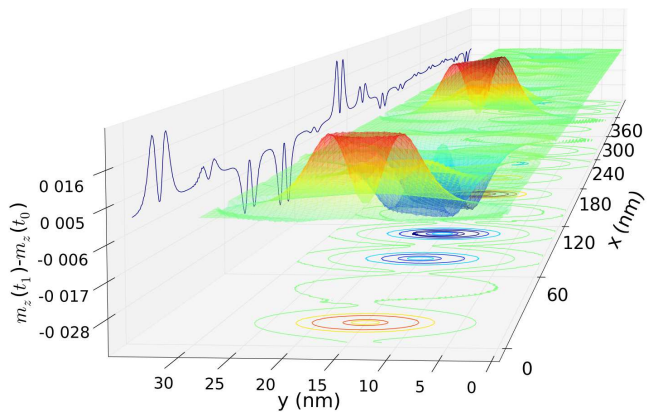}
\caption{(Color online) The $m_{z}(t_{1})-m_{z}(t_{0})$ of collective breathing mode excited with the frequency $f_{\mathrm{signal}}=12.984$ GHz.}
\label{3D}
\end{figure}

By extending system to the chain of closely spaced 20 nanodisks, we create the 1D magnonic crystal based on skyrmions. In Fig. \ref{structure}(b) the distribution of $m_z(t_0)$ in the array consisting of 20 nanodisk of 30 nm diameter and 1 nm thickness is presented. The periodic boundary conditions are additionally implemented within mumax$^3$ with finite number of the repetitions (15 in our simulations) along the $x$ direction to serve for minimization effects of the demagnetizing fields from the external edges of the array.\cite{4899186}  

In order to obtain dispersion relation of dynamical skyrmions a 2D Fourier transform (2DFT) (from space and time domain $(x,t)$ to wavenumber and frequency domain $(k,f)$) is performed on  magnetization $z$ component, $m_{z}(x,y,t)$ at value of $y$ corresponding to center of the dots. The calculated dispersion relation of dynamical skyrmions in the array of nanodots is shown in Fig.~\ref{disp}. This dispersion was obtained with out-of-plane excitation field $h_{\mathrm{dyn}}$, applied locally to the disk in center of the chain (disk no. 9). Two  magnonic bands are found in frequencies below $f_{\mathrm{signal}}$, both have clear periodicity (with the period $G$) in dependence on the wavenumber and Brillouin zone (BZ) boundary at $k=\pi/a \approx 0.1$ nm$^{-1}$. The finite size of the structure used in simulations is seen as discrete states in the dispersion relation. This discrete character might be eliminated by increasing the number of states (increasing number of disks), "smearing" of discrate states (increasing damping factor value) or by influencing the strenght of interactions and distance between discrate states (by changing geometry of the structure).

Frequencies of magnonic bands in Fig.~\ref{disp} (0.49-0.99 GHz and 11.88-13.68 GHz) correspond to frequencies of the modes in the isolated dot. Thus, it suggests that these magnonic bands can be related to clockwise gyrotropic mode and breathing mode branch, respectively. In order to determine character of the modes and to visualize the spatial distribution of the dynamic magnetization component $\delta m_{z}$, we follow the procedure described in Ref.~[\onlinecite{kumar2013effect}]. We perform 2DFT on $m_{z}$ for each value of $y$ seperately. In the next steps we define a 1D vector consisting of values of obtained 2DFT at choosen frequency, we choose the wavenumber $k$, keep the values of 2DFT at all wavenumbers with the value differing by reciprocal lattice vector $\textbf{G}= 2\pi /a$: $k\pm n G$, where $n$ is integer and $a$ lattice constant (in our case $a=$ 30 nm) and set the zero at other wavenumbers. Finally we perform inverse Fourier transformation, resulting in dynamic magnetization component for mode with specific frequency and wavenumber $\delta m_{z}$. The results are presented in Fig.~\ref{profiles} (a) and (b). The spatial distribution of $\mathrm{Re}(\delta m_{z})$ in magnonic crystal clearly exhibit similarities with $\mathrm{Re}(\delta m_{z})$ and $\mathrm{Im}(\delta m_z)$ of isolated dots (Fig. \ref{single} (a) and (b)).

The band width of the breathing mode branch is more than three times wider than the width of the clockwise gyrotropic mode branch. This points at the much stronger coupling between breathing oscillations. Moreover, the group velocities have opposite signs in these two branches. Those features point at the different coupling mechanisms responsible for propagation in each case. We showed, that the breathing mode has dominating oscillation of the $m_{z}$ component of the magnetization (see Fig. \ref{single} (b)). Thus, the coupling between skyrmion excitations in nanodots is realized mainly via out-of-plane components of $\mathbf{M}$. The lower frequency of collective skyrmion is expected for antiphase oscillation of $m_{z}$ in the neighbor nanodots (assuming prevailing role of $m_z$ in coupling), analogically to a chain of parallel magnetic dipoles oriented perpendicularly to the chain axis, where the state of lower energy is for antiferromagnetic arrangement of dipoles. In magnonic crystals, such oscillations are realized at the BZ boundary (where wavelength of the excitation is equal $2a$), while in the BZ center all nanodots oscillate in-phase. This explain the frequency decrease of the breathing mode with increasing wavenumber from zero to BZ boundary, i.e., negative group velocity.  
For the gyrotropic mode the interaction is governed by the effective net magnetization induced by the core shift, similar to the gyrotropic modes in nanodots in the vortex state.\cite{99} 

Another interesting point is that in the case of isolated disk we have found in simulations only one mode (breathing) when the out-of-plane magnetic field was applied, whereas in case of the array of dots, two branches (breathing and clockwise branch) are excited with out-of-plane dynamic magnetic field applied to the single dot. We suppose that it is due to generation of demagnetizing field with three nonzero magnetization components. It means that the demagnetizing magnetic field exerted by the oscillations in the excited dot couple to the low frequency gyrotropic mode in the near nanodots. Approximatly 10 times smaller intensity of the breathing band dispersion then the gyrotropic mode band found in calculations (Fig. \ref{disp}) confirms this hypothesis. The further investigations are necessary to identify if the process of coupling is between breathing oscillations and gyrotropic modes or only between forced oscillation in single dot and gyrotropic motion.

The interesting property learned from the dispersion relation in Fig.~\ref{disp} is also possibility for dynamic magnetization transmission through the skyrmion array. This is due to non-zero group velocity for $k \neq n \pi/a$. We excite skyrmion mode with sinusoidal microwave magnetic field oscillating at $f_{\mathrm{signal}}=12.984$ GHz (which correspond to  $k= 0.4 (\frac{\pi}{a}$)) and plot $m_{z}(t_{1})-m_{z}(t_{0})$ at selected time $t_{1}=1.5 \times 10^{-9}$ s in Fig.~\ref{3D}. We apply the external magnetic field only at the first disk. The results presented in Fig.~\ref{3D} confirm propagation of the breathing skyrmion mode. 

In summary, we have shown that magnonic band structure based on skyrmion excitations are possible to form in array of nanodisks with DMI. The dispersive character and collective properties of the excitation were demonstrated with micromagnetic simulations. Analogically to collective vortex excitation, they are potentially applicable for logic devices \cite{103,jung2012logic} and can be verified experimentally with time-resolved scanning transmission x-ray microscopy STXM \cite{99}. The research opens a new direction in research of skyrmions and might be extended to 2D or even 3D artificial structures and also any aperiodic structure in order to obtain desired functionality. Also the tailoring of dispersion propertis of dynamical skyrmions might be realized by skyrmion phase control in neighbouring dots, e.g., antiparallel aligment. A promissing applications of the propesed 1D magnonic crystals and collective excitations are related to their combining with single dot excitation of the skyrmion dynamics by spin torque nanoscillators.\cite{zhang2015current}  

\section*{Acknowledgements}
The research leading to these results has received funding from Polish National Science Centre project DEC-2-12/07/E/ST3/00538. The numerical calculation were performed at Poznan Supercomputing and Networking Center (grant No 209). 
 
\bibliographystyle{apsrev4-1}
\bibliography{books}

\end{document}